\documentclass{aa}

\usepackage{lipsum} 
\usepackage[utf8]{inputenc}
\usepackage{amsmath}
\usepackage{natbib}
\usepackage{graphicx}
\usepackage{comment}

\usepackage{txfonts}

\usepackage{hyperref}
\hypersetup{colorlinks,citecolor=blue,linkcolor=blue}

\begin{document}

   \title{Flux ropes and dynamics of the heliospheric current sheet}

   \subtitle{Study of a PSP / Solar Orbiter conjunction with 3D MHD simulations}

    \author{V. Réville\inst{1}
    \and N. Fargette\inst{1}
	\and A.P. Rouillard\inst{1}
	\and B. Lavraud\inst{2,1}
	\and M. Velli\inst{3}
	\and A. Strugarek\inst{4}
	\and S. Parenti\inst{5}
	\and A.S. Brun\inst{4}
	\and C. Shi\inst{3}
	\and A. Kouloumvakos\inst{1}
	\and N. Poirier\inst{1}
	\and R.F. Pinto\inst{4}
	\and P. Louarn\inst{1}
	\and A. Fedorov\inst{1}
	\and C.J. Owen\inst{6}
	\and V. Génot\inst{1}
	\and T.S. Horbury\inst{7}
	\and R. Laker\inst{7}
	\and H. O'Brien\inst{7}
	\and V. Angelini\inst{7}
	\and E. Fauchon-Jones\inst{7}
	\and J.C. Kasper\inst{8}} 
    \institute{IRAP, Université Toulouse III - Paul Sabatier, CNRS, CNES, Toulouse, France
    	\and Laboratoire d'astrophysique de Bordeaux, Univ. Bordeaux, CNRS, Pessac, France
		\and UCLA Earth, Planetary and Space Sciences, CA, USA
		\and D\'epartement d'Astrophysique/AIM, CEA/IRFU, CNRS/INSU, Univ. Paris-Saclay \& Univ. de Paris, 91191 Gif-sur-Yvette, France
		\and Université Paris-Saclay, CNRS, Institut d’Astrophysique Spatiale, 91405, Orsay, France
		\and Mullard Space Science Laboratory, University College London, Holmbury St. Mary, Dorking, Surrey, RH5 6NT, UK.
		\and Imperial College London, South Kensington Campus, London, SW7 2AZ, UK
		\and BWX Technologies, Inc, Washington DC 20001 USA}

  \abstract
   {Solar Orbiter and PSP jointly observed the solar wind for the first time in June 2020, capturing data from very different solar wind streams, calm and Alfvénic wind as well as many dynamic structures.}
   {The aim here is to understand the origin and characteristics of the highly dynamic solar wind observed by the two probes, in particular in the vicinity of the heliospheric current sheet (HCS).}
   {We analyse the plasma data obtained by PSP and Solar Orbiter in situ during the month of June 2020. We use the Alfvén-wave turbulence MHD solar wind model WindPredict-AW, and perform two 3D simulations based on ADAPT solar magnetograms for this period.}
   {We show that the dynamic regions measured by both spacecraft are pervaded with flux ropes close to the HCS. These flux ropes are also present in the simulations, forming at the tip of helmet streamers, i.e. at the base of the heliospheric current sheet. The formation mechanism involves a pressure driven instability followed by a fast tearing reconnection process, consistent with the picture of \citet{Reville2020ApJS}. We further characterize the 3D spatial structure of helmet streamer born flux ropes, which seems, in the simulations, to be related to the network of quasi-separatrices.}
   {}


   \maketitle
%

\section{Introduction}

\label{intro}

The launch of Solar Orbiter on February 10, 2020 \citep{Muller2013,Muller2020}, has opened very promising opportunities for multi-spacecraft in situ observations of the inner heliosphere in conjunction with the Parker Solar Probe (PSP) \citep{Velli2020} and other spacecraft such as BepiColombo and the Solar-Terrestrial Relations Observatory (STEREO). The first measurements of Solar Orbiter were collected in the spring of 2020 during a very interesting time window that coincided with PSP's fifth solar encounter.  Multipoint in-situ measurements are essential to help us understand the structure and dynamics of the inner heliosphere and, as we shall illustrate, they are ideally complemented by 3D MHD simulations. Simpler models such as the potential source surface model \citep[PFSS,][]{AltschulerNewkirk1969,Schatten1969}, although very useful, do not describe the plasma properties measured by the probes, in particular the differences between steady and more dynamic solar wind states. Indeed, the degree of variability of the solar wind may be an important marker of its coronal origins  \citep[see, e.g.][]{Antiochos2012}. Here, variability refers to fluctuations in speed, density and magnetic field on all scales, and is different from the omnipresent turbulence that is also an important part of the more steady wind components, coming from coronal holes. Turbulence in the latter states is dominated by Alfv\'enic fluctuations, with correlations corresponding to propagation away from the Sun, within streams of typical speeds greater than 600 km/s \citep[see, e.g.][and references therein]{TuMarsch1995}. However, there are also slow wind streams of high Alfvénicity, as discussed in \citet{DAmicis2021}, and PSP has observed many such streams in the inner heliosphere. In contrast, what we refer to as the intrinsically dynamic solar wind component appears to be mostly slow, and its fluctuations do not show strongly Alfvénic correlations. The origin of this slow wind appears to lie in proximity of closed coronal structures, such as the system of loops encountered in bipolar streamers or pseudo-streamers \citep[][]{Antiochos2011}.

For more than twenty years, density structures propagating from the low corona into the solar wind have been observed with white-light coronagraphs and heliospheric imagers \citep[see, e.g.][]{Sheeley1997,Wang1998,DeForest2018}. These structures display some periodicity \citep{Viall2009,Viall2010}, and their internal magnetic structure is likely to consist of flux ropes \citep{SanchezDiaz2017a,SanchezDiaz2019}, i.e., helical magnetic field structures that often arise from reconnection events. The HCS, although very thin, is surrounded by a thicker, denser layer: the heliospheric plasma sheet (HPS). \citet{Lavraud2020} have analyzed recent observations of PSP, arguing that the high-beta plasma of the HPS consists mainly of material expelled from a reconnection process. Numerical simulations have indeed shown that the HCS is unstable and that reconnection occurs at the tip of helmet streamers \citep{Einaudi1999,Endeve2003,Endeve2004,Rappazzo2005,HigginsonLynch2018}. In \citet{Reville2020ApJL}, using high-resolution resistive 2.5D MHD simulations, we provided evidence that a fast tearing instability \citep{Furth1963,Biskamp1986,Loureiro2007,PucciVelli2014,Tenerani2015} was the process responsible for the release of plasmoid-like structures and density perturbations. 

The global picture in 3D is, however, more complex. What processes control reconnection at the base of a warped HCS? Moreover, realistic solar magnetic field can develop other substructures that can favor the accumulation of currents and reconnection. \citet{PriestDemoulin1995,Demoulin1996} have developed the concept of quasi-separatrix layers (QSLs), where the connectivity of magnetic field lines, though remaining continuous, experiences large gradients. QSLs can be identified computing the squashing factor $Q$, which quantifies the gradients of connectivity of field lines between two arbitrary surfaces \citep{TitovDemoulin1999,Titov2002,Titov2007}. These layers have been further shown to develop currents and trigger 3D reconnection \citep[see][]{Aulanier2005,Aulanier2006} in MHD simulations in the context of solar flares. QSLs have been also used to explain the presence of a slow and dynamic wind at high latitudes, away from the HCS, through reconnection triggered by footpoint motions in the very low corona and propagating along the fan of pseudo-streamers \citep{Antiochos2011,Higginson2017}.

This paper aims at identifying and characterizing the sources of slow solar wind variability and dynamics observed close to the current sheet by PSP and Solar Orbiter. In section \ref{sec:overview}, we detail the context of observations from both probes during the month of June 2020 and the procedure adopted to select the solar magnetograms used as inputs for the 3D MHD simulations. In section \ref{sec:numerics}, we briefly present the MHD model and compare the global outputs of two simulations with the in situ measurements of PSP and Solar Orbiter.  Section \ref{sec:FR} presents our main results. We first characterize the flux rope events in Solar Orbiter and PSP data, using magnetic field as well in-situ measurements of ions and electrons. We note a good correspondence between the flux rope events in the data and their occurrence in the simulations. We thus study the properties of flux ropes generated in the simulations. We show that their periodicity is consistent with the process described in \citet{Reville2020ApJL}, given the numerical constraints of 3D simulations. Finally, we discuss for the first time the longitudinal structure and distribution of 3D flux ropes along the HCS, and show that it could be related to the network of QSLs. We summarize and discuss these results in section \ref{sec:ccl}.

\section{Overview of the period}
\label{sec:overview}
\subsection{Available data and spacecraft positions}

\begin{figure}
\center
\includegraphics[width=3.3in]{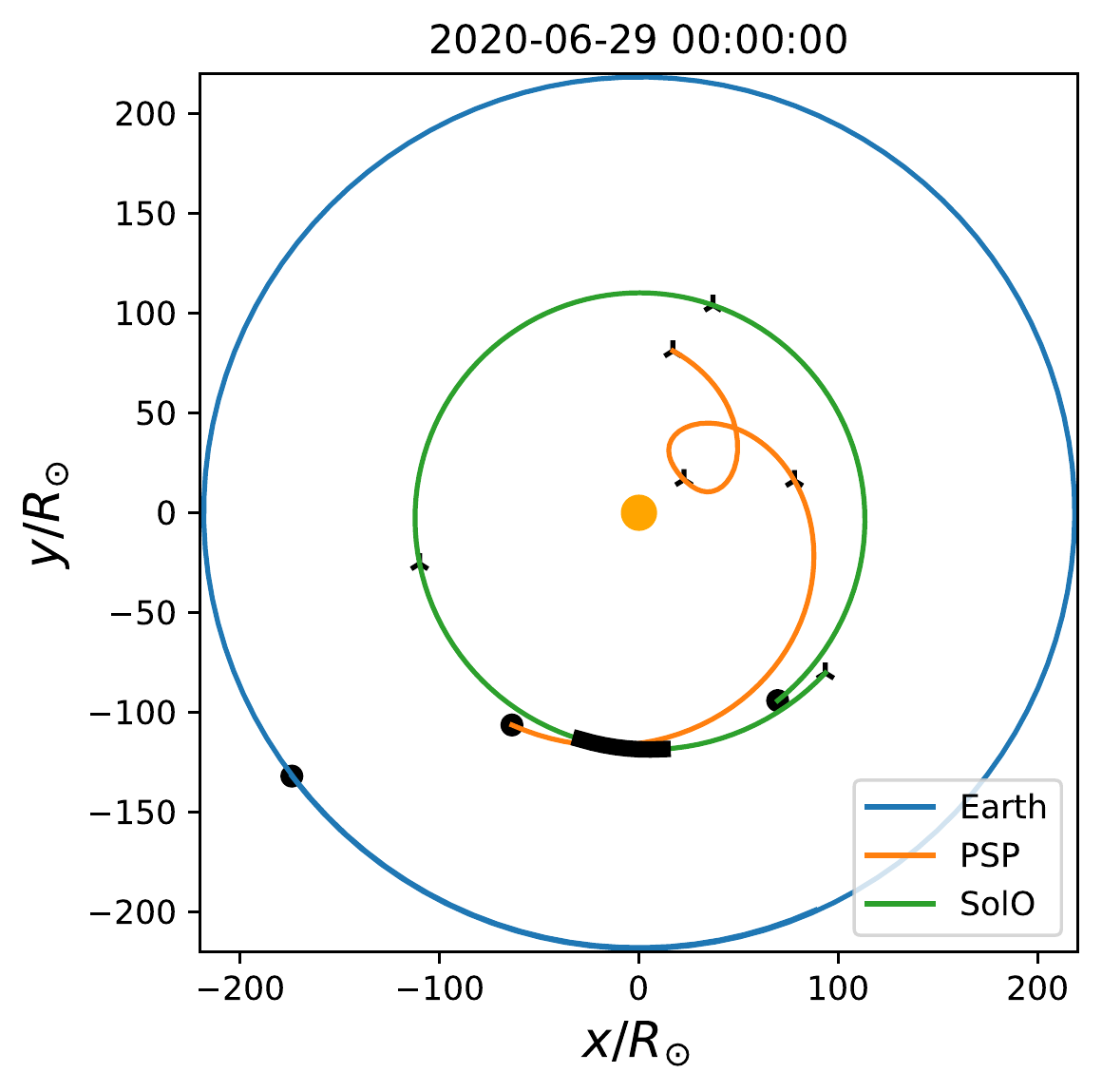}
\caption{Relative positions of PSP and Solar Orbiter in the solar Carrington frame on June 29, 2020. The back traced trajectory goes back to May 27 and ticks indicates the positions of the spacecraft every 11 days (the whole interval represents 33 days). The thick black line corresponds to available measurements of SWA/PAS.}
\label{fig:context}
\end{figure}

PSP went through its fifth encounter during the month of June 2020 with its perihelion at $27.8R_{\odot}$ on June 7, 08:00 UTC. The first perihelion of Solar Orbiter occurred on June 15 08:00 UTC, with the closest approach at $111 R_{\odot}$, about $0.5 AU$. Figure \ref{fig:context} shows the positions of the spacecraft, in the Carrington rotating frame, on June 29, 2020, as well as each spacecraft's previous 33 days of orbital trajectory. The thick black line region shows the region where SWA/PAS \citep[the Solar Wind Analyser Proton and Alpha Sensor, see][]{Owen2020} was able to record measurements of the full 3D velocity distributions of ions, between May 30 and June 1. Both spacecraft took measurements of the magnetic field vector during the entire encounters with the FIELDS \citep[onboard PSP,][]{Bale2016} and MAG \citep[onboard Solar Orbiter,][]{Horbury2020} instruments. PSP also made plasma measurements with the SWEAP/SPC (Solar Probe Cup), SWEAP/SPAN-i (electrostatic analyzer for ions) and SWEAP/SPAN-e (electrostatic analyzer for electrons) over the whole month \citep{Kasper2016}. PSP's distance to the Sun varied between  $27.8R_{\odot}$ and $125 R_{\odot}$, while Solar Orbiter remained around $0.5$ AU over the whole period. We see in Figure \ref{fig:context} that Solar Orbiter has made a full rotation over the Sun in the Carrington frame and thus has taken measurements at all Carrington longitudes. As for latitudes, PSP was between -3.9 and 2.7 degrees, Solar Orbiter between 3.3 and 6.6 degrees. Note that the black region in Figure \ref{fig:context} has been probed by the two spacecraft about a month apart, at the end of May by Solar Orbiter, and at the end of June by PSP. 

\subsection{Choice of the magnetograms}

\begin{figure*}
\center
\includegraphics[width=7in]{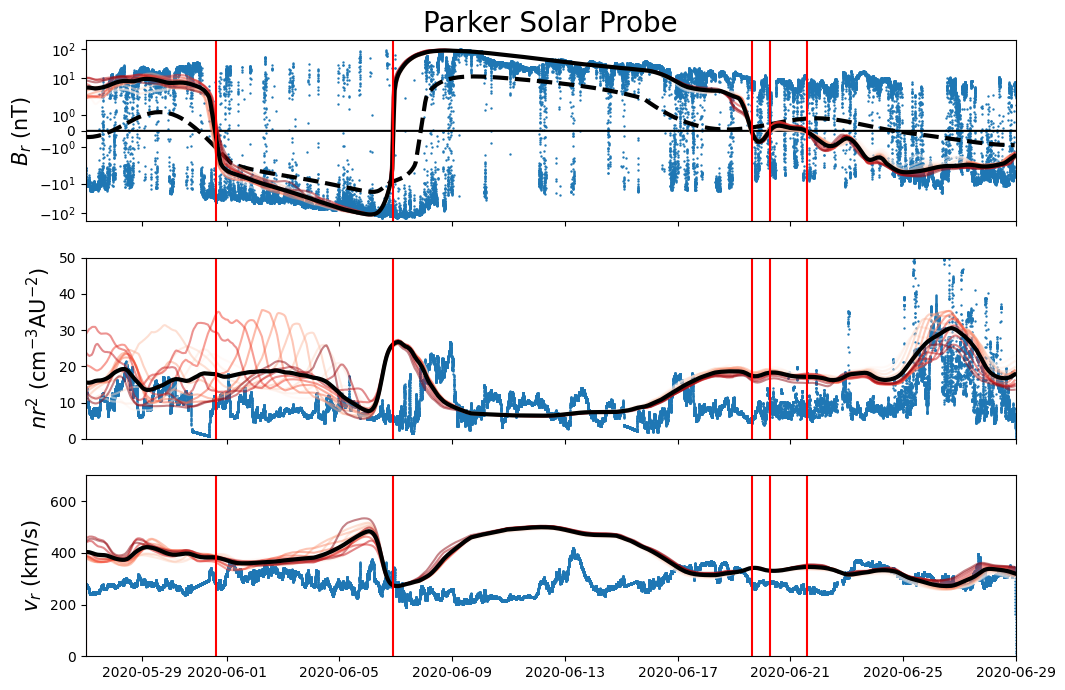}
\caption{In situ measurement taken by PSP between May 27 and June 29 of 2020. Magnetic field measurements are made with FIELDS, while the plasma properties are obtained with SWEAP/SPC and are shown in blue. The results of the two simulations are shown in shades of red for 13 outputs separated by two hours, while the average is shown in black. The red vertical lines are the polarity changes observed in the simulations. The PFSS solution for the radial field obtained in section \ref{sec:overview} is shown in dashed black in the top panel.}
\label{fig:psp_insitu}
\end{figure*}

In this study, we use for the initialization and for the magnetic boundary conditions of the MHD simulation, global magnetograms of the photospheric magnetic flux, computed by the Air Force Data Assimilative Photospheric Flux Transport (ADAPT) model \citep{Arge2010, Arge2013}. The ADAPT maps are global magnetograms produced using data assimilation techniques based on National Solar Observatory Global Oscillations Network Group (NSO/GONG) measurements, along with a magnetic flux transport model. They provide different realizations of the photospheric magnetic field at a certain time, which are the result of the evolution of the magnetic field by the flux transport model on the far side of the Sun. In \citet{Reville2020ApJS}, we chose to rely on an ADAPT magnetogram, taken on November 6, 2018, for the whole 45-day period of the first perihelion. Some mismatch has been observed, but the overall behavior of the connection could be predicted. For this period of June 2020, connectivity studies have proved to be more difficult, possibly due to the emergence and disappearance of active regions on the solar disk \citep[see][for a detailed study on the shape of the HCS during this period]{Laker2021}. To prepare the MHD simulations, we thus perform a preliminary study of the sector boundaries predicted by a potential field source surface (PFSS) model \citep{Schatten1969,AltschulerNewkirk1969,Badman2020ApJS}. We wish to select the minimum amount of synoptic maps that best reproduce the polarity of the magnetic field measured by the two probes, and consequently the right source for the solar wind.

We optimize the time spent within the correct polarity assuming a given magnetic map for both spacecraft, i.e. we looked for the maximum of:
\begin{equation}
\begin{split}
    f(M) =& \frac{1}{2} \int_{t_1}^{t_2} \left(\mathrm{sign}(B_{r,PSP}(t)  B_{r,M}(t))  + 1\right) \\
	&+ \left(\mathrm{sign}(B_{r,SO}(t)  B_{r,M}(t))+1\right) dt,
\end{split}
\label{eq:pfss_opt}
\end{equation}
where $B_{r,PSP}$ and $B_{r,SO}$ are the radial field measurements obtained by the two spacecraft and $B_{r,M}$ is the radial field interpolated on the projection of each trajectory on the source surface radius assuming a Parker Spiral at a given wind speed. The integrated function in equation \ref{eq:pfss_opt} equals to $1$ when both spacecraft $B_r$ measurements agree with the PFSS solution, $1/2$ for only one spacecraft, and $0$ otherwise. We search for the magnetic map that maximizes $f(M)$ in all ADAPT realizations at 00:00 and 12:00 each day between May 27 and June 30 of 2020. We set, as fixed parameters, the source surface radius at $2 R_{\odot}$, and the wind speed at $300$ km/s, which is close to the averaged speed observed by both spacecraft during the period (see section \ref{subsec:insitu}).

\begin{figure*}
\center
\includegraphics[width=7in]{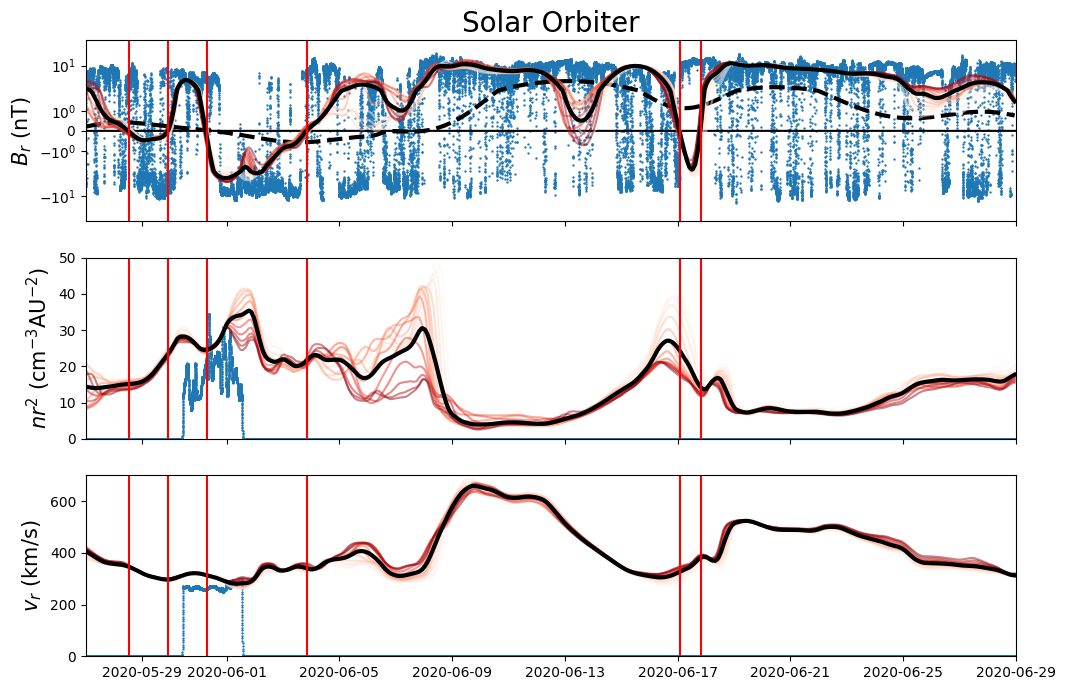}
\caption{In situ measurement taken by Solar Orbiter between May 27 and June 29 of 2020. Magnetic field measurements are made with MAG, while the plasma properties are obtained with SWA/PAS (only two days of measurements). The results of the plasma measurements interpolated in the simulations are shown as in Figure \ref{fig:psp_insitu}. The PFSS solution for the radial field obtained in section \ref{sec:overview} is shown in dashed black in the top panel.}
\label{fig:slo_insitu}
\end{figure*}

We find that it is reasonable to split the month of June in two time intervals, with a given optimum for both spacecraft on each interval. We first select a period between May 27 and June 6. We find that the second realization ADAPT map of June 1 at 00:00 UTC best fits the polarity of both spacecraft before June 6. For the second period, from June 6 to June 30t, the ADAPT map of June 14 at 12:00 UTC (realization 2) gives the best score with equation \ref{eq:pfss_opt}. The need for a minimum of two time intervals is related to the appearance of an active region on June 3 on the solar east limb. In Figure \ref{fig:psp_insitu} and Figure \ref{fig:slo_insitu}, we show, in the upper panels, the results of this preliminary study. The radial field measurements of both spacecraft are shown in blue, while the results of the PFSS models in dashed black lines. 

\section{MHD modelling and global results}
\label{sec:numerics}

\subsection{Alfvén wave driven MHD model}
We use the Alfvén wave driven MHD model of \citet{Reville2020ApJS}, which we call WindPredict-AW, to unravel the global heliospheric structure at the time of the measurements. The full set of equations and boundary conditions can be found in the latter paper, though some further improvements have been brought to the model. First, as in \citep{Reville2020ApJL}, the non-solenoidal condition on the magnetic field is ensured through the constrained transport method \citep{Dedner2002}. This feature is actually important when dealing with increasingly complex magnetic field associated with field aligned thermal conduction. The divergence cleaning method used in \citet{Reville2020ApJS} can create oscillations in the field direction, and subsequent numerical instabilities due to the behavior of the thermal conduction in small coronal loops (in regard to the resolution) typical of active regions. Hence, as the solar cycle goes to a more active phase, we found this update to be a very useful feature of the PLUTO code \citep{Mignone2007}, on which WindPredict-AW is based. 

Second, the model domain now includes the transition region. The inner boundary condition lies in the chromosphere with temperature of $2 \times 10^4$ K. The density is set to a value of $2 \times 10^{10}$ cm$^{-3}$. We implemented the technique described in \citet{Lionello2009} to thicken the transition region in order to ease the numerical computation and limit the numerical resolution needed. The transverse velocity perturbation parameter, which controls the amplitude of the Alfvén waves launched in the domain, is set to $\delta v = 12$ km/s. This is lower than the value used in \citet{Reville2020ApJS}, as Alfvén waves are strongly amplified in the transition region. The input Alfvén wave pointing flux is $F_p = \rho_\odot v_A \delta v^2 \sim 1.5 \times 10^5$ erg.cm$^{-2}$.s$^{-1}$. For this study, we use uniform grids in the angular direction, with $160$ and $320$ cells in the $\theta$ and $\phi$ direction, i.e. a $1.125$ degree resolution. In the radial direction, the grid is separated in three regions. The first region, meant to render the modified transition region, has 10 cells between $1.0$ and $1.004 R_{\odot}$. The grid is then stretched up to $15R_{\odot}$, with a minimal resolution of $0.5 R_{\odot}$, and stretched again up to $130 R_{\odot}$ with the largest radial grid size of $4 R_{\odot}$. The simulations are performed in the rotating Carrington frame with a sidereal period of $25.38$ days for a duration of $176$ hours (physical time), on the Jean-Zay super computer (IDRIS/CNRS France). The two simulations represent roughly 800k CPU hours. 

\subsection{Comparison with in situ data}
\label{subsec:insitu}

Let us now compare the results of the MHD simulation with the data of PSP and Solar Orbiter. Figure \ref{fig:psp_insitu} shows three panels with the radial magnetic field, the radial velocity and the particle density. The magnetic field is shown in symmetrical logarithmic scale to better cope with the variation of PSP's distance to the Sun. Similarly, the density is normalized to the radial distance squared, expressed in astronomical units. The data is shown in blue. We show, at one minute resolution, magnetic field measurements from FIELDS as well as the radial proton speed and density derived from the L3 moments of the SWEAP/SPC measurements. PSP's coordinates are retrieved from the open-source python package \textit{heliopy} \citep{HeliopyStansby2021} and converted to the Carrington frame of the simulations. Then, we interpolate the simulations' fields along the spacecraft trajectories. We superpose the results of several outputs of the simulations in red shading, spaced by 2 hours between $t=132h$ and $t=158h$, and show the average in black. The transition between the first and the second simulation is done over 6h centered around June 6, 2020 12:00 UTC. The red vertical lines correspond to the heliospheric current sheet crossings detected in the simulation, corresponding to sign changes of the $B_r$ average profile.

Looking solely at the black averaged line, one notices an overall agreement for the main large-scale fluid moments between the simulations and the data. On PSP's solar approach, from May 27 to the perihelion of June 7, we observe one main HCS crossing which is well reproduced by the simulation using the map of June 1. The amplitude of the radial magnetic field appears fully consistent with the data, especially away from the current sheet, which is however less sharp in the simulation than in reality. Around the time of this HCS crossing, we observe strong perturbations in the density as well as smaller ones in the radial velocity. They are due to a very large scale flux rope propagating in the simulation. Close to PSP's perihelion, there is another HCS crossing, which occurs around June 6 22:00 UTC in the simulation, roughly one day before the data. Despite this delay, the structure of the HCS crossing in the magnetic field and in the density is well reproduced by the simulation. An episode of very calm wind then follows, up to June 18. Again, the magnetic field and the density are close to the averaged data. The model predicts in general a faster wind speed, especially for this quiet wind interval, where the simulation speed is around 500 km/s, against an average of 300 km/s in the data. The data shows nonetheless a velocity peak of 400 km/s around June 14. 

Starting roughly after June 17, the wind comes back to a denser, slower and variable state both in the data and in the simulation (based on the June 14 map). As we shall discuss in the next section, PSP measured a wind pervaded by multiple HCS crossings and small to average size flux ropes. This variability is also present in the simulation, as shown by the deviation from the average of the red curves in the density, velocity and magnetic field. Finally, late in the month (June 27), we observe a full crossing of the HCS associated with an increased in density. The crossing is not reproduced by the simulation, but the density increase is, which means that the simulated HCS is close to PSP's coordinates at these times.

We repeat the same operation in Figure \ref{fig:slo_insitu} with the data of Solar Orbiter. The magnetic field is again displayed in symmetrical logarithmic scale, cadenced at one minute, and comes from the L3 public data of the MAG instrument. The density and the velocity panels display the data obtained with the proton and alpha sensor onboard Solar Orbiter. The SWA/PAS measurements represent about two days of data and include one current-sheet crossing, which is well reproduced by the simulation. Between May 27 and June 9, Solar Orbiter crossed the HCS many times and measured a lot of variability, which is consistent to the variability present in the simulations. Later on, the wind is faster and calmer, except for two briefs magnetic polarity changes simulated on June 13th and June 17, which seem to correspond to multiple HCS crossings measured in situ at these times.

\section{HCS dynamics and flux ropes}
\label{sec:FR}

\subsection{Flux rope characterization in Solar Orbiter and PSP data}

\begin{figure}
\includegraphics[width=3.5in]{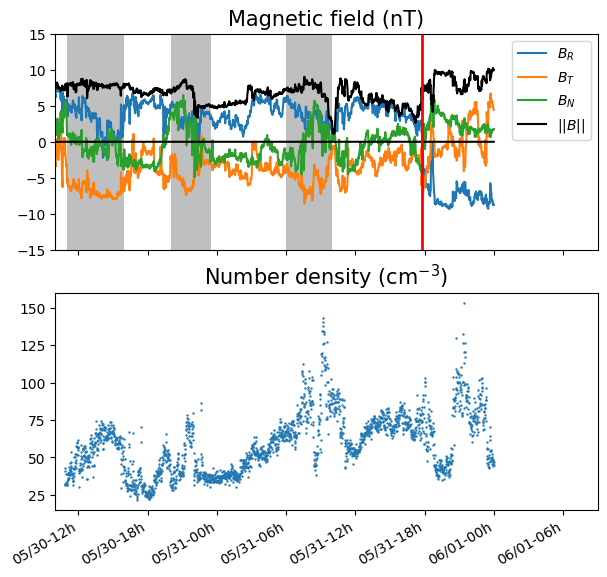}
\caption{Probable flux rope events (in gray) observed by Solar Orbiter on May 30 and 31 before crossing the HCS (in red). All events are characterized by an increase of the total field strength and density, and a bipolar structure in one component of the magnetic field (here $B_N$).} 
\label{fig:fr_slo}
\end{figure}

In this section, we further study the dynamics and variability observed in the data and the simulations. Despite the very short time window of PAS measurements, we have been able to identify in Solar Orbiter's data a series of probable flux-rope signatures on May 30 and May 31. In Figure \ref{fig:fr_slo}, we show the magnetic-field components in the $RTN$ frame as well as PAS' proton density measurements. We observe characteristic features of flux ropes: increase of the total field strength, changes in plasma density and temperature, and a sign change of one of the magnetic field components. Solar Orbiter was close to the HCS at the time of these events and crossed it a few hours later (red line in Figure \ref{fig:fr_slo}). The marked flux rope periods show a bipolar structure in $B_N$, which is the normal component to the spacecraft trajectory, but is also likely to be close to the normal to the HCS. They also show a strong tangential component $B_T$, which is consistent with a configuration where the flux rope is mostly in the plane of the HCS. 

\begin{figure}
\includegraphics[width=3.5in]{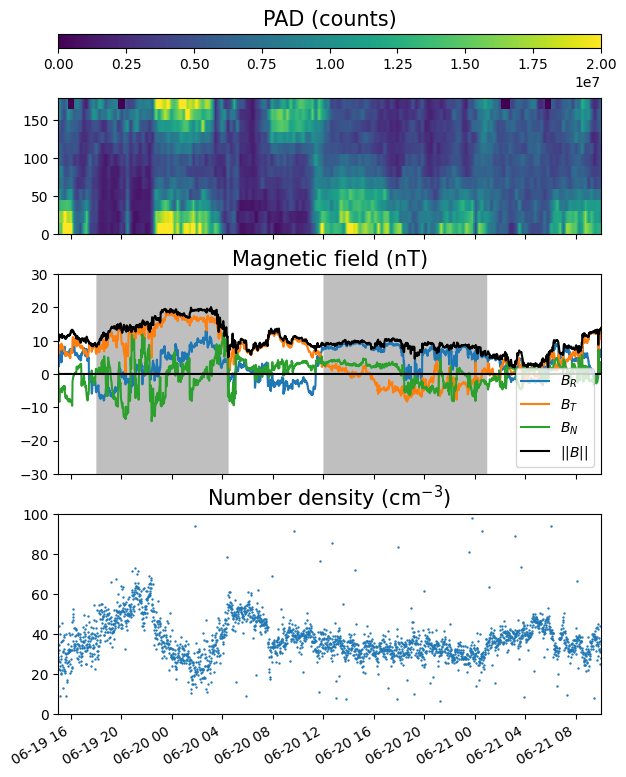}
\caption{Two flux ropes structure identified in PSP data. The top panel shows the PAD from SPAN-e, in arbitrary count units, of electrons between 283.9 and 352.9 eV. The middle and bottom panel shows the structure of the magnetic field in the $RTN$ frame and the plasma density (SWEAP/SPC). The two gray shaded area indicate the position of flux ropes with different connectivity visible in the PAD, uni-directional (right) and bidirectional electrons (left).} 
\label{fig:fr_psp}
\end{figure}
As shown in Figure \ref{fig:context}, the same region of the inner heliosphere was probed by PSP about a month later and found to exhibit similar features. In Figure \ref{fig:fr_psp}, we show the structure of two flux rope events measured by PSP near June 20. We add to the previous analysis the electron pitch-angle distribution (PAD) shown in the top panel, and obtained with SWEAP/SPAN-e. We focus on the 8th channel of SPAN-e, which captures electrons with energy between 283.9 and 352.9 eV, corresponding to the suprathermal strahl electron population at this distance from the Sun \citep[see, e.g.][]{Gosling2005}. The first event is on the left of the panel (starting on June 19 at 18:00), where we observe an increase in the total magnetic field strength, as well as a full reversal of the radial and normal field. The pitch-angle distribution shows electrons fluxes aligned and anti-aligned to the magnetic field, suggesting that we are in a structure connected to the Sun at both ends. The second event occurs right after a HCS crossing and lasts from June 20 at 12:00 to June 21 at 1:00 UTC. We see a smaller increase of the magnetic field amplitude, with a reversal of $B_N$. The plasma density drops somewhat, while the PAD show electron fluxes parallel to the magnetic field, which means that the structure is connected to the Sun at one end only. Right after the flux rope (in gray), we also observe a strahl electron drop out, characteristic of a reconnection region disconnected from the Sun \citep{Gosling2005}.

\begin{figure*}
\center
\includegraphics[width=7in]{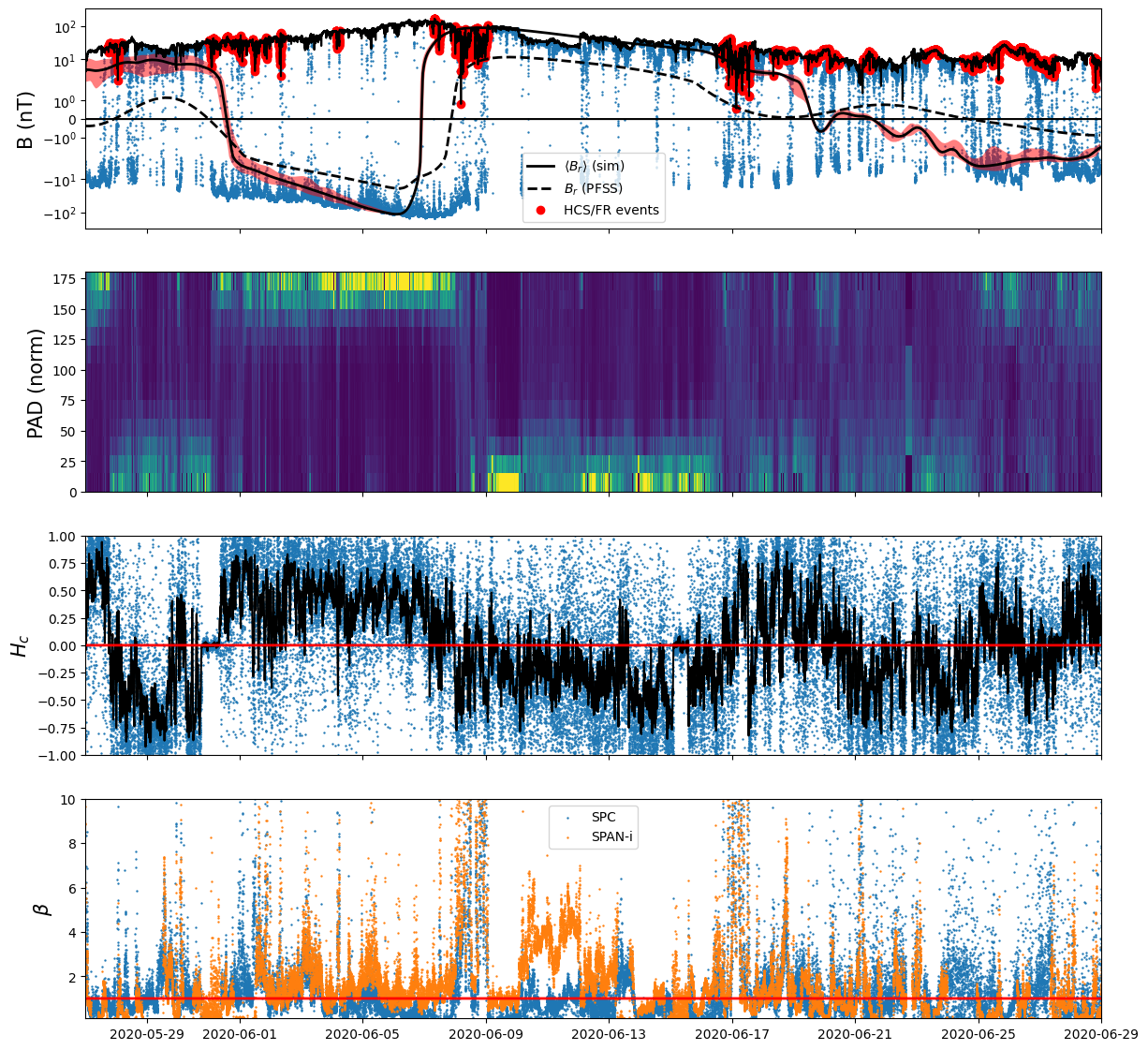}
\caption{PSP measurements of dynamical structures. In the first panel, we show the observed radial (blue) and total (black) magnetic field over the month of June. The results of the simulation and of the PFSS models are repeated from Figure \ref{fig:psp_insitu}, and the variability of the simulations is shown in shaded red. Red points identify flux rope events and partial or total HCS crossings observed in the data on the total magnetic field line. They are further identified with the second panel, showing the normalized pitch-angle distribution of electrons in SPAN-e 8th channel, around 300eV. The third panel shows the cross-helicity of perturbations in the velocity and magnetic field, and is close to one in negative polarity, close to minus one in positive polarity of the interplanetary magnetic field. The bottom panel shows the value of the plasma beta parameter obtained with SPC and SPAN-i, and the red line indicate a value of $\beta =1$.} 
\label{fig:beta_comp}
\end{figure*}

We now consider the whole month of PSP measurements. A joint analysis of the PAD, magnetic field, density and velocity field is performed to identify HCS crossings and flux using four different instruments, FIELDS, SWEAP/SPC, SWEAP/SPAN-i and SWEAP/SPAN-e, and summarized in Figure \ref{fig:beta_comp}. The first panel resembles the first panel of Figure \ref{fig:psp_insitu}, but shows additional information. The variability of the radial field in the simulations is shown in shaded red, representing the min and the max value of the curves shown in Figure \ref{fig:psp_insitu}. The total measured magnetic field is plotted in black, overlaid in red is the occurrence of flux rope and/or HCS (partial) crossings. Appendix \ref{app:FRtable} lists and labels all the events represented in red, with precise start and stop times. The PAD is normalized to the energy integral for a given time, as the electron counts vary greatly with heliocentric radial distance. The third panel shows the cross-helicity of the solar wind perturbations defined by : 
\begin{equation}
    H_c = \frac{2 \delta \mathbf{v} \cdot \delta \mathbf{b}/\sqrt{4\pi \langle \rho \rangle}}{\delta v^2 + \delta b^2/(4 \pi \langle \rho \rangle)},
\end{equation}
with
\begin{eqnarray}
    \delta \mathbf{v} &= \mathbf{v} - \langle \mathbf{v} \rangle,\\
    \delta \mathbf{b} &= \mathbf{B} - \langle \mathbf{B} \rangle,
\end{eqnarray}
and the averaging operator $\langle \rangle$ representing a running average of one hour. It is interesting to notice the good correspondence between the pitch-angle distribution and the cross-helicity defined this way. Because electrons are always streaming away from the Sun, when the interplanetary magnetic field is globally Sunward, the PAD distribution is mainly concentrated around 180 degrees. In such situation, for a relatively calm wind, Alfvén waves are expected to propagate away from the Sun too and the correlation between the velocity and magnetic field perturbations is positive, yielding a cross-helicity up to one. This correspondence works well for the beginning of the interval up to June 17. We also observe clear changes in the pitch-angle distribution of suprathermal electrons (going from 0 to 180 degrees or the opposite), which are indicators of HCS crossings, for instance on May 29 and on June 8 right after PSP's perihelion. 

The last panel of Figure \ref{fig:beta_comp}, shows the plasma $\beta$ parameter obtained with SPC and SPAN-i. $\beta$ covers a wide range of values, sometimes below one, but clear enhancements are observed either close to the HCS crossing or during flux-rope events. After June 17, the PAD shows fast variations of streaming electron directions, as well as strahl electron dropouts. The cross-helicity also suddenly rises when the polarity of the magnetic field stays globally positive. This period between June 17 and June 25 is a very good example of what happens when PSP cruises very close to the HCS. In addition to multiple crossings, many flux ropes are observed, which is interestingly very consistent with the variability seen in the simulations, as shown in Figure \ref{fig:psp_insitu}. 

\subsection{Origin of simulated flux ropes}

\begin{figure*}
\begin{tabular}{cc}
\includegraphics[width=3.5in]{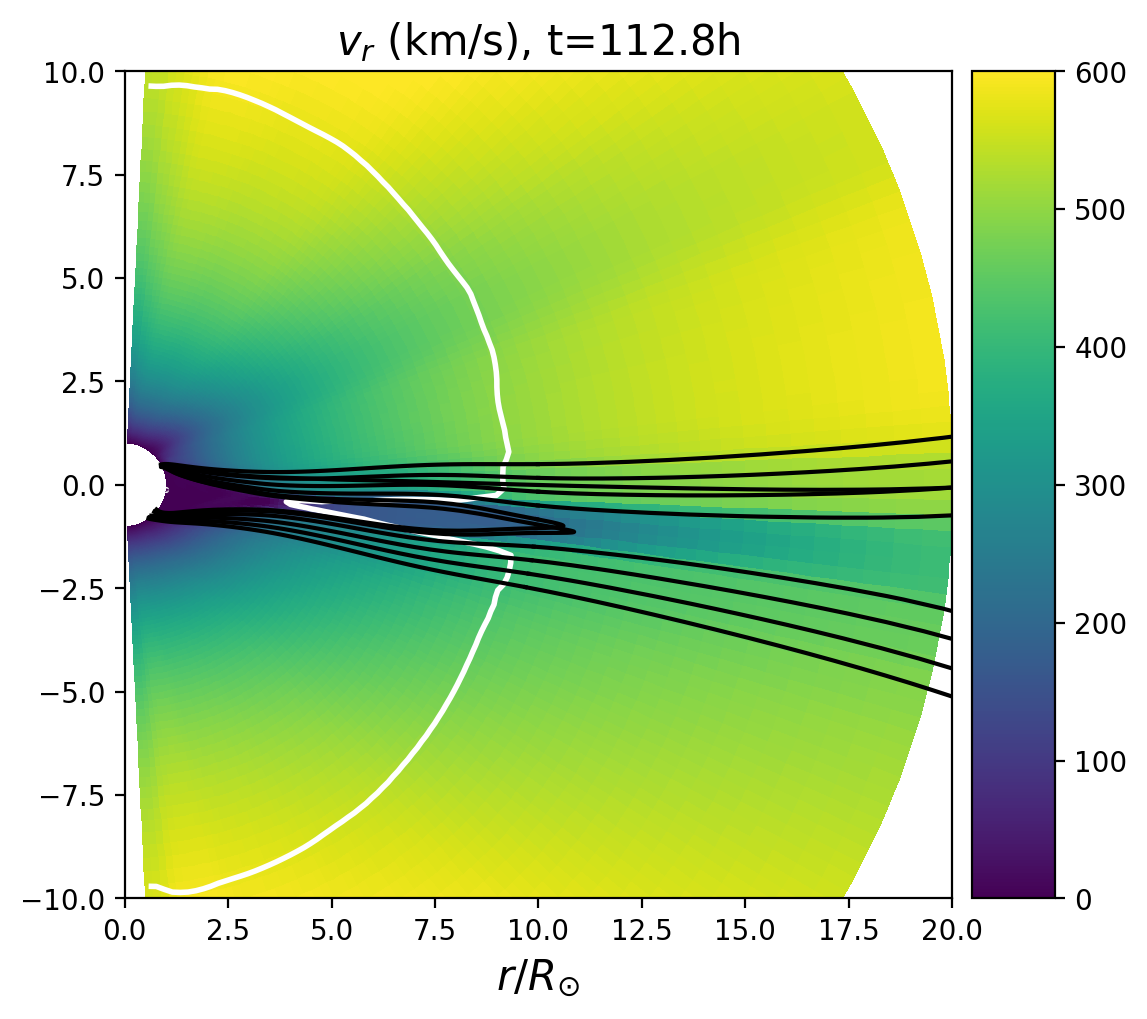} & \includegraphics[width=3.5in]{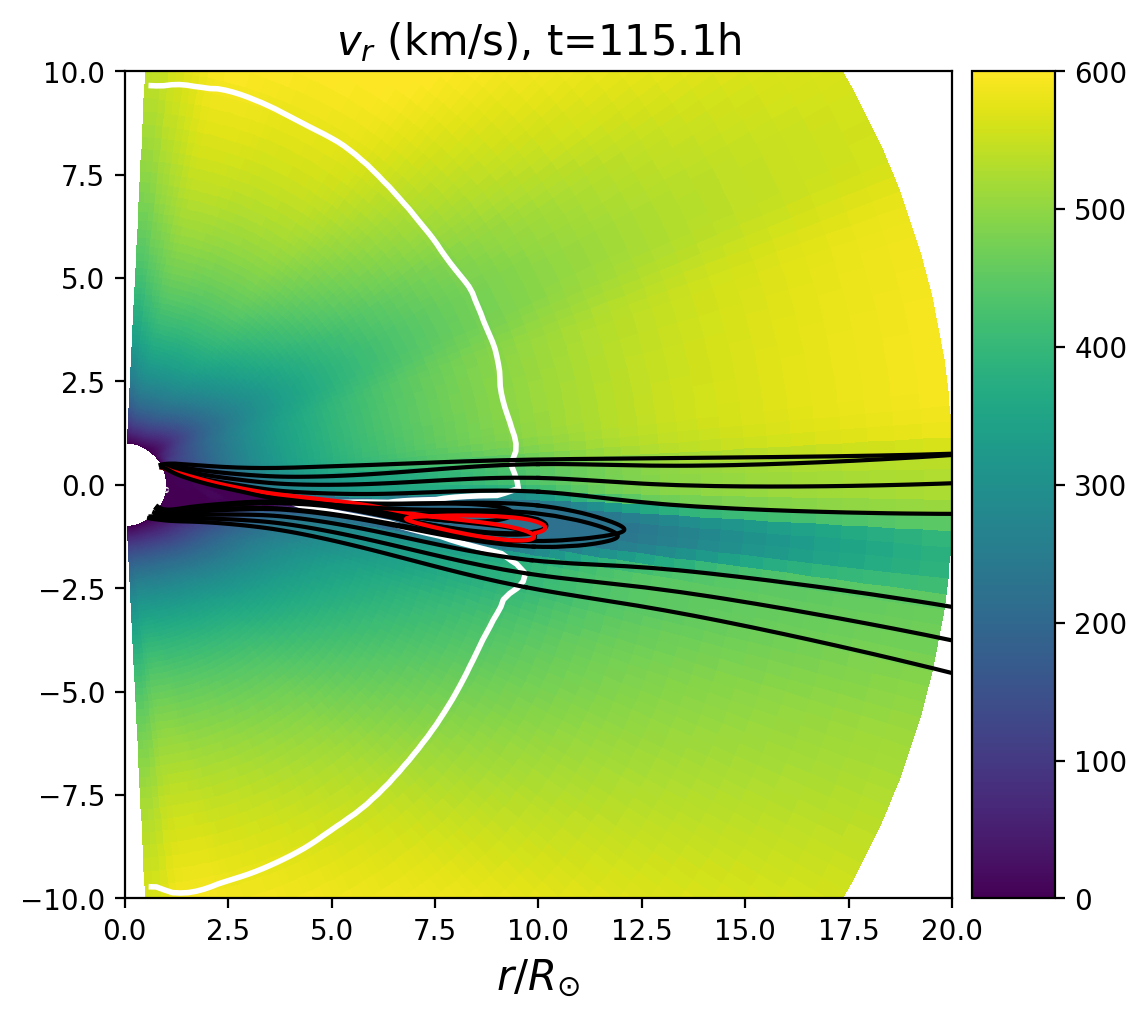}\\
\includegraphics[width=3.5in]{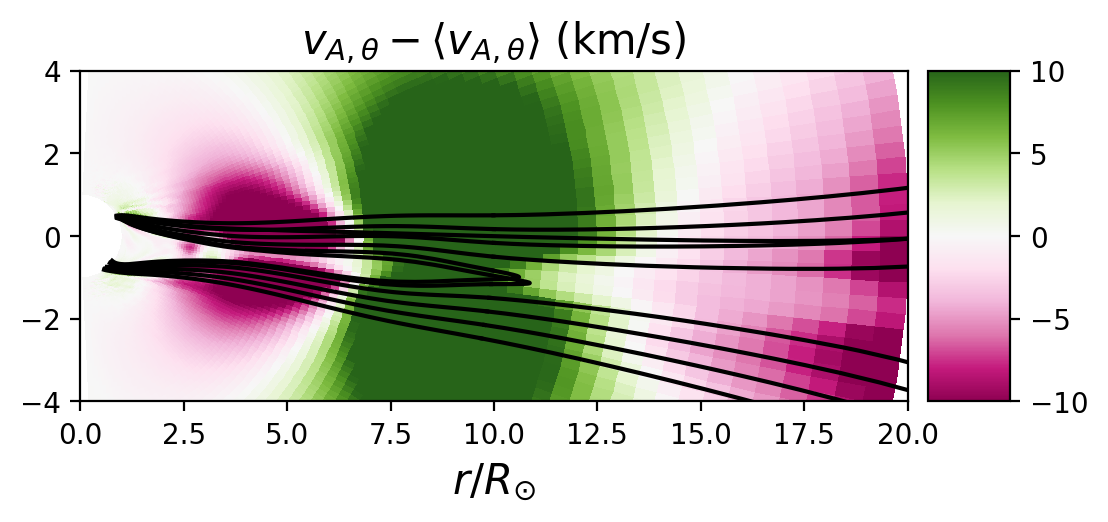} &
\includegraphics[width=3.5in]{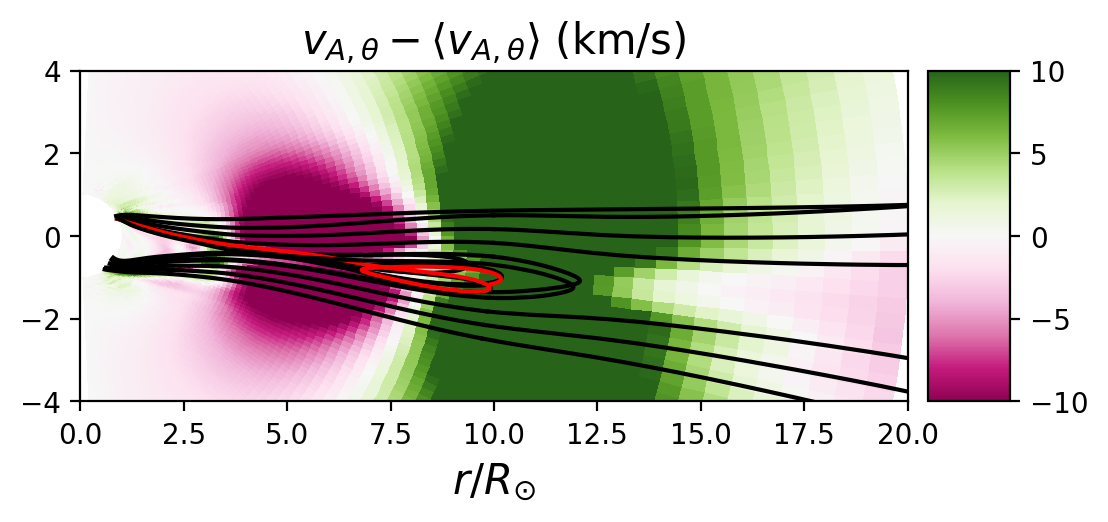} \\
\includegraphics[width=3.5in]{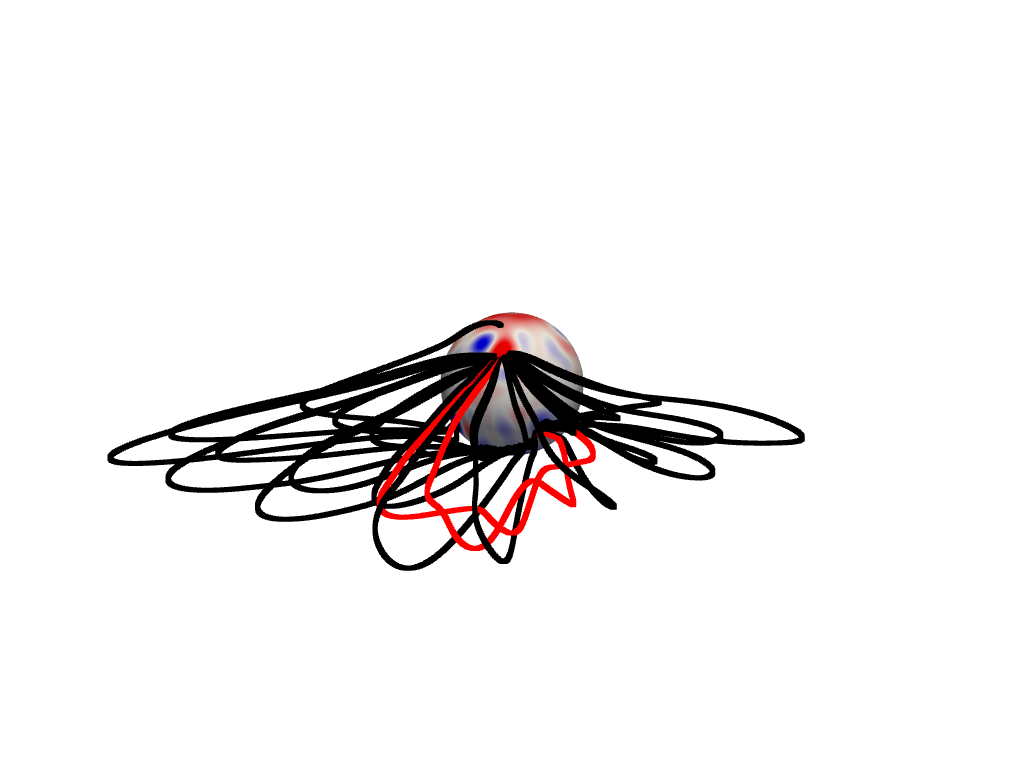} & \includegraphics[width=3.5in]{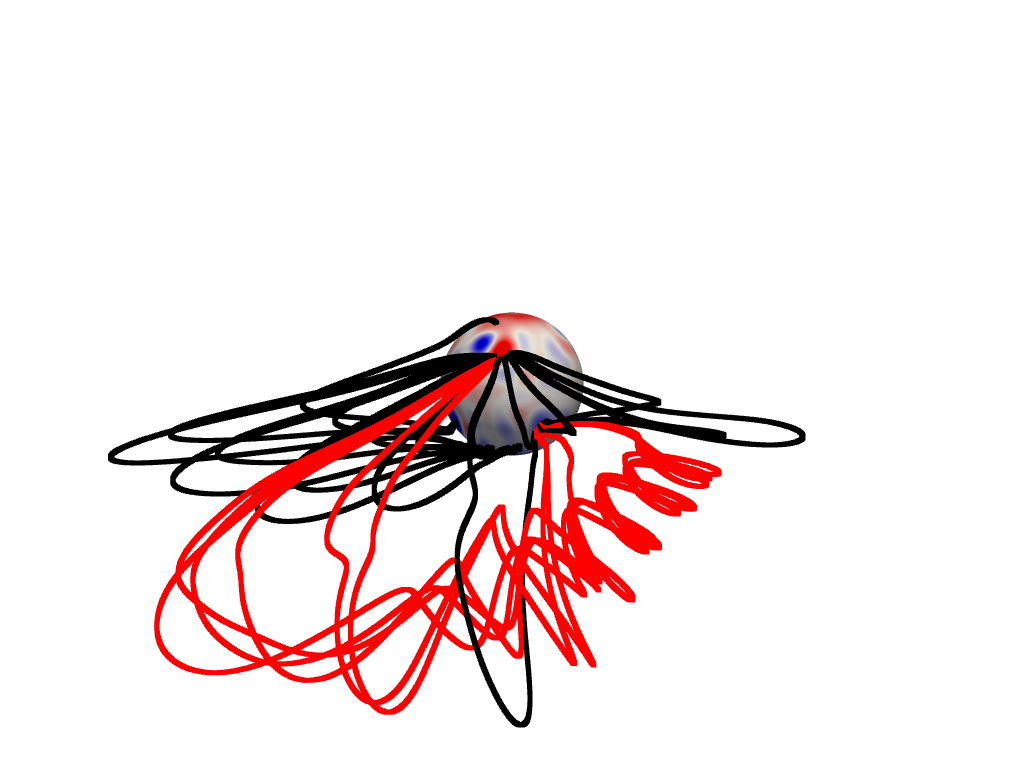}
\end{tabular}
\caption{Onset of the reconnection in the June 14 simulation. The top panels show a 2D view of field lines traced from the HCS and connecting to the surface at two epochs, $t=112.8h$ and $t=115.1h$. The view is a cut at $\varphi=\pi$ (180° longitude). The background color shows the velocity field. The white rounded curve shows the Alfvén surface. Field lines are plotted in red when they possess helical structures. The middle panel shows the perturbations in the tangential field (in units of Alfvén velocity). A large-scale tearing mode can be clearly identified at the reconnection locus. The bottom panel shows a 3D view of the same epoch before and after the main reconnection event and the creation of the flux rope.}
\label{fig:FR3D}
\end{figure*}

What is the process responsible for the creation of these flux ropes? In our simulations, after roughly 60h of simulated time, reconnection develops in the current sheet between 5 and 10 solar radii. Following this reconnection, helical structures --flux ropes-- are created and propagate in the solar wind. In Figure \ref{fig:FR3D}, we show snapshots of the June 14 simulations at two different times. On the left panel, at $t=112.8h$, we show a 2D cut at $\varphi=\pi$ (180° longitude) of the corona and solar wind velocity along with selected field lines around the current sheet. The middle panel displays the perturbed tangential Alfvén speed for the same time interval. The average $\langle v_{A,\theta} \rangle$ is computed over an interval of 20h around the reconnection time. Above the main streamer, we observe a growing long wavelength mode characteristic of the tearing instability. This perturbation is propagating and is about (at $t=112.8h$) to fully cross the Alfvén surface (in white). In the bottom panel, we render a 3D view of the magnetic-field geometry. We observe a tilt of the field lines to the left, sign of the presence of a tangential field oriented along $-\mathbf{e}_{\varphi}$. Field lines in red have started the reconnection process (note that they are not exactly corresponding to the 2D view). On the right of Figure \ref{fig:FR3D}, we show the same features, at $t=115.1h$. The HCS has reconnected and created a flux rope following the tangential or "core" field orientation. Magnetic footpoints of the flux rope have considerably shifted to the right and the structure extends to about 50 degrees of longitudes. We can see the markers of the acceleration following the reconnection, with field lines advected further away on the left than on the right.

\begin{figure}
\includegraphics[width=3.5in]{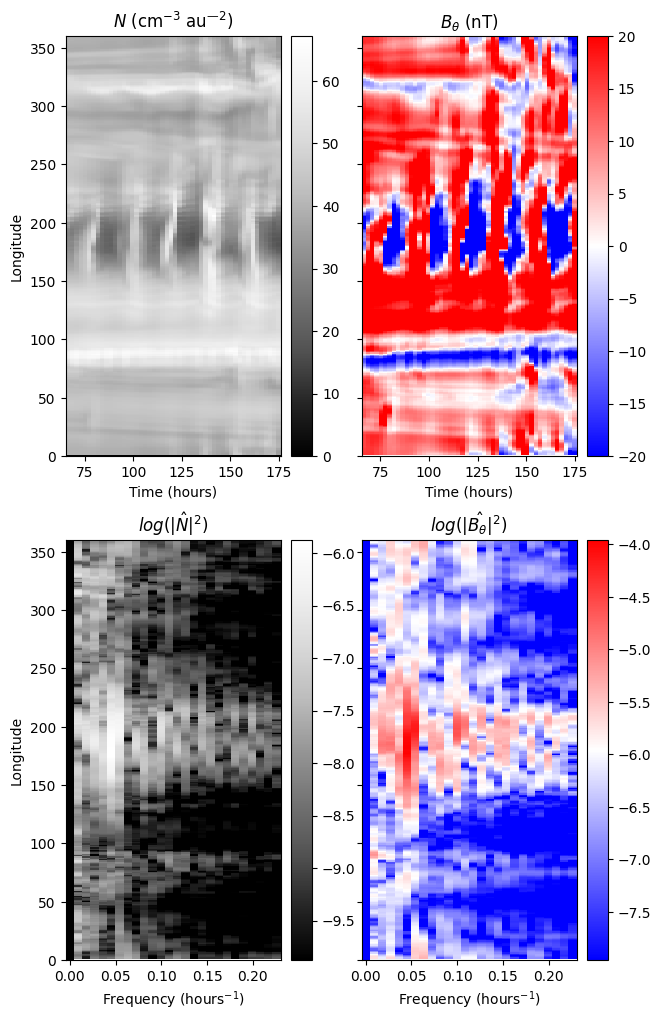}
\caption{Variability of the HCS at $15 R_{\odot}$ for the June 14 simulation. In the top panel, the number density $N$ and the latitudinal magnetic field $B_{\theta}$ are extracted along the HCS and stacked with time. Several propagating flux ropes are going through the domain, with various longitudinal extent, but a rather steady time pattern. In the bottom panel, we perform a Fourier analysis in time (for each longitude). For all quantities, the peak is located around a period of 20h.}
\label{fig:Fourier}
\end{figure}

In Figure \ref{fig:Fourier}, various 3D quantities are extracted along the HCS at $15 R_{\odot}$ for the June 14 simulation. Signatures of the reconnection process can be seen easily in the latitudinal magnetic field. Density enhancements are located on the front of the flux ropes, consistently with the picture given in \citet{SanchezDiaz2017a,SanchezDiaz2019}. Flux ropes are more present at certain longitudes. Figure \ref{fig:Fourier} shows a rather extended perturbed front between 130 and 300 degrees of longitudes. This front is likely made of several structures released periodically. Another smaller flux-rope front is observed around a longitude of 360 at the top of the figures. To characterize the periodicity of these multiple flux ropes, we plot in the bottom panel of Figure \ref{fig:Fourier} the temporal Fourier spectra of all four quantities computed for each longitude. For the main flux rope structure, there is a clear peak at a frequency of $0.05$ hours$^{-1}$, or a periodicity of 20 hours. This periodicity is also seen in the smaller structure at longitudes around 360 (it is especially visible in the spectral plot of $B_{\theta}$). The process at the origin of these periodic release is thus relatively independent of the longitude and of the magnetic structure of the corona. In \citet{Reville2020ApJL}, we identified two characteristic periods. The longer periodicity \citep[around 30h in][]{Reville2020ApJL}, was proposed to be associated with the formation and re-formation of the tip of the helmet streamers, whose timescales are related to the coronal heating process of the simulation. We find this longer periodicity here with the 3D simulations, a periodicity that is also observed in the solar wind \citep[e.g.][]{Viall2015,SanchezDiaz2017a}, as well as other works on streamers' instability \citep{Endeve2003,HigginsonLynch2018}. 

\begin{figure}
    \centering
    \includegraphics[width=3.5in]{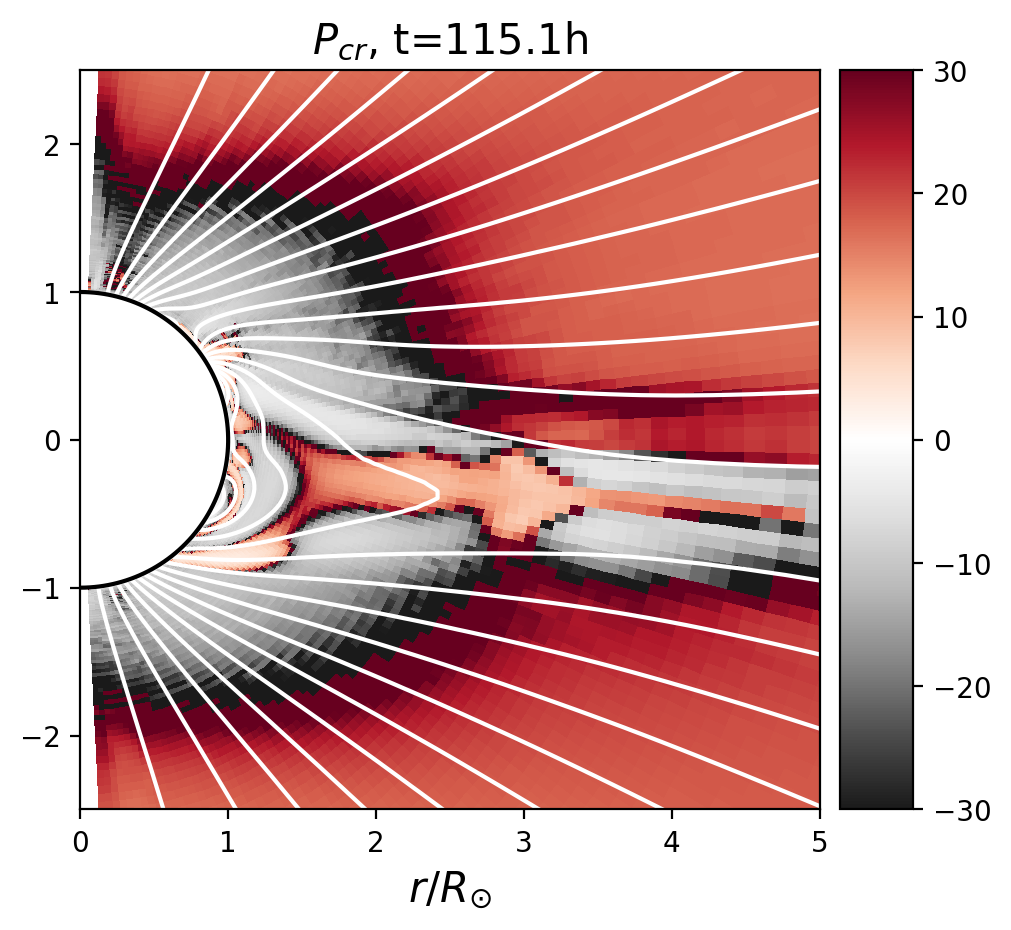}
    \caption{Period of the pressure driven instability in the close corona. The panel shows a meridional cut at $\phi = \pi$ ($180^\circ$ longitude) for the June 14 simulation. Grey regions mark stable zones. In the inner corona, we clearly identify the tip of helmet streamers and the HCS as unstable zones, with typical $P_{\mathrm{cr}} \sim 10-30$ hours.}
    \label{fig:stab_period}
\end{figure}

This 10 to 30h long period can be understood as the result of a pressure driven instability followed by a tearing of the current sheet, leading to magnetic reconnection. Simply put, at the tip of helmet streamers, the pressure gradients are sufficient to transport plasma parcels against the magnetic tension of the closed loop. This then results in an elongation and thinning of the current sheet that becomes unstable to a tearing mode. This process has been described experimentally by \citet{Peterson2021} and we follow a modified version of the heuristics proposed in this work, to compute the typical timescale of the lengthening of the current sheet. Defining $\mathbf{\kappa} = (\mathbf{B} \cdot \nabla) \mathbf{B}/||B||^2$, the magnetic field curvature vector, $p=p_{th}+p_w$ the sum of the thermal and wave pressure, and $\mathbf{g}$ the acceleration of the Sun's gravity, the transverse displacement $\xi$ for a parcel of plasma can be written \citep[neglecting terms in 1/$\beta$, see][]{Peterson2021}:

\begin{equation}
    \frac{1}{R_c} \frac{\partial^2 \xi}{\partial t^2} = \mathbf{\kappa} \cdot (\frac{c_s^2}{p} \nabla p-\mathbf{g}),
    \label{eq:taucr}
\end{equation}
where $R_c=1/||\mathbf{\kappa}||$, is the curvature radius of the magnetic field. When the right-hand side of equation \ref{eq:taucr} is negative, the magnetic tension can confine the plasma inside the helmet streamer. When it is positive, the plasma displaced according to the solution of equation \ref{eq:taucr}:
\begin{equation}
    \xi(t)  = R_c \left[\kappa \cdot (c_s^2 \nabla p/p - \mathbf{g})\right] t^2/2.
\end{equation}
As shown in Figure \ref{fig:FR3D}, the plasma parcel is displaced typically out to $10R_{\odot}$, while the tip of helmet streamers is located around $2R_{\odot}$, hence $\xi \sim 8 R_{\odot}$. The periodicity of the whole process can thus be computed as:
\begin{equation}
    P_{cr}= \mbox{sign} \left[\kappa \cdot (c_s^2 \nabla p /p - \mathbf{g})\right] \sqrt{\frac{2 \xi}{R_c \mathbf{\kappa} \cdot (c_s^2 \nabla p/p-\mathbf{g})}},
    \label{eq:pcr}
\end{equation}
where we kept the sign function to characterize stable regions $(P_{\mathrm{cr}} < 0)$. Figure \ref{fig:stab_period} shows a 2D colormap of $P_{\mathrm{cr}}$. Computing the average period inside unstable regions of a shell between $r=1.3R_{\odot}$ and $r=2R_{\odot}$, we get $\langle P_{cr} \rangle \sim 19$ hours, which is very close to the typical period obtained in the simulations. Note that the boundaries between stable and unstable zone might play a significant role in this average value, which makes sense as these boundaries will be the first displaced during the process.

This process shares many similarities with our previous work in 2.5D. In \citet{Reville2020ApJL}, we proposed that the main process responsible for reconnection at the current sheet is a tearing mode. The time to thin the current sheet and trigger reconnection was between $60$h and $90$h and increasing with $S=Lv_A/\eta$, the Lundquist number. A tearing mode was then created, with plasmoids ejected beyond the Alfvén critical point in the slow solar wind. The typical scale of the plasmoids is a decreasing function of the Lundquist number, and we found, for low S regime, a wavelength of about $2R_{\odot}$, which is roughly what we observe here in 3D. In the 2.5D study, the Lundquist number had to be higher than a critical value of $\sim 10^4$ for the tearing mode to be triggered. In the present 3D simulations, we do not include explicit resistivity and as such, it is the numerical resistivity that acts for the reconnection process. Because of computational costs, it is much harder to reach the necessary resolution for a proper description of the tearing process. Using a simple dimensional analysis, we can estimate $\eta = V \Delta L$, where $V$ is a typical wave speed of the problem, say the Alfvén speed and $\Delta L$ the grid resolution. The numerical Lundquist number $S=Lv_A/\eta$ then reduces to $S_{\mathrm{num}}=L/\Delta L$, and considering the reconnection region ($r=5R_{\odot}$) and the resolution across the current sheet we get $S_{\mathrm{num}} \sim 500-1000$, which is below the threshold observed in 2.5D resistive MHD. However, it is also possible that in 3D the onset of the tearing mode is lowered due to the presence of the core field and/or complex magnetic fields coming from the magnetic map. One important effect is an ``effective" finite extent of the x-line in the third dimension that might allow a disconnection of a flux rope even at low S.  Nevertheless, the high S, asymptotic behavior, characterized by an extremely thin sheet with the formation of plasmoid chains and secondary tearing \citep[seen in][]{Reville2020ApJL}, is absent in the 3D simulation and further research is necessary to better characterize this process in 3D.

\begin{figure}[!ht]
\includegraphics[width=3.5in]{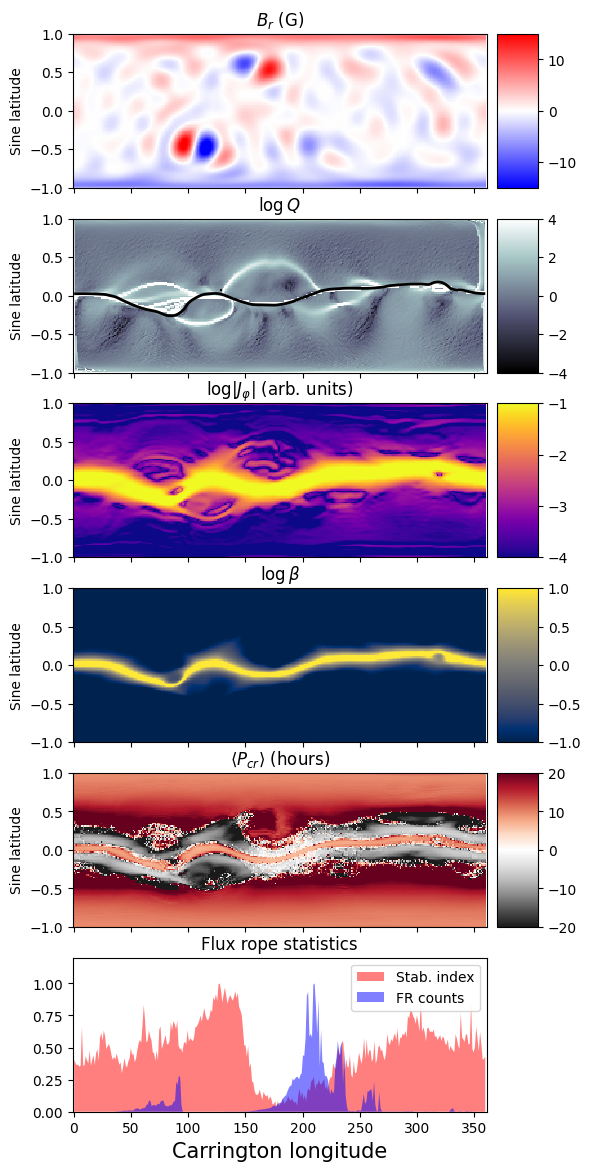}
\caption{Spatial structure of the June 14 simulation. The top panel shows the surface magnetic field obtained from the ADAPT magnetogram. The second panel displays the logarithm of the squashing factor $Q$. The maximum of $Q$ matches the HCS, shown in black, while secondary arches indicates QSLs. The third panel shows the logarithm of the azimuthal current $J_{\varphi}$ at $3 R_{\odot}$. Maxima are located at the HCS with some variation of intensity along the sheet. Black patches indicate where the current is above 70\% of the maximum. Currents also appear at quasi-separatrices. The fourth shows the logarithm of the $\beta$ parameter at $3 R_{\odot}$. Higher $\beta$ are observed at the HCS and close to the HCS/QSLs intersections. The average $\langle P_{cr}(r=3R_{\odot}) \rangle$ is shown in the fifth panel, displaying consistently stable regions at the QSL/HCS intersections. Finally, the last panel shows the normalized histogram of detected flux ropes in the simulation in blue and the stability index described in the text.}
\label{fig:squashing}
\end{figure}

3D simulations can bring information on the distribution and the longitudinal and latitudinal extent of these flux ropes. 2.5D plasmoids should generalize to full torii in 3D, but this is obviously not what we observe here in our simulations (see Figure \ref{fig:FR3D}) and more generally in the solar wind \citep{Rouillard2008,Rouillard2010}. Simulated flux ropes are confined in latitude, extending by a few degrees above and below the HCS. In the simulations, flux ropes show different kinds of connectivity. As shown in Figure \ref{fig:FR3D}, they start connected to the solar surface by both legs, which would correspond to observed bidirectional electrons in Figure \ref{fig:fr_psp}. But we have also observed configurations where the flux rope is connected on one end to the Sun and the other one into the solar wind, corresponding to the uni-directional electron flux shown in Figure \ref{fig:fr_psp}.

In Figure \ref{fig:squashing}, we analyze the 3D longitudinal distribution of these flux ropes. In the top panel, we show the structure of the surface magnetic field imposed as the boundary condition in the simulation of June 14. We observe notably two strong bipolar structures, around 100$^\circ$ and 150$^\circ$ longitude, that correspond to two active regions present at the time. The 3D structure of the MHD solution is obviously largely shaped by the structure of the surface magnetic field. However, features in the solar wind -i.e. beyond closed regions extending to a few $R_{\odot}$ are more easily interpreted computing the network of separatrices and quasi-separatrices. In the second panel of Figure \ref{fig:squashing}, we plot the squashing factor $Q$ in logarithmic scale as a longitude/latitude synoptic map. $Q$ is computed between two spherical surfaces at $r=r_{\odot}$ and $r=5 R_{\odot}$ following the formulation given in \citet{Titov2007}. The HCS, defined as $B_r(\theta, \varphi) = 0$, is shown in black and corresponds to the maximum of the current. The squashing factor identifies high gradients and discontinuities in the connectivity of field lines, and thus either true separatrices like the HCS where $Q$ should be in principle infinite, or QSLs for high values ($Q \geq 10^3$). QSLs form arched structures that correspond in general to the location of pseudo-streamer fans. The two bipoles around 100$^\circ$ longitude create pseudo-streamers and the largest structures in the network of QSLs. Their signatures can be seen in most other MHD variables.

In the second panel, we show a cut of the logarithm of $|J_{\varphi}|$ at $r=3 R_{\odot}$. Although most of the current structures are located around the HCS, weaker currents are present along the QSLs. Magnetic field shears can indeed be easily created there following small perturbations \citep[see][]{Aulanier2005,Aulanier2006}. The structure of quasi-separatrices is also visible in $\beta$, plotted in the fourth panel of Figure \ref{fig:squashing}. At the HCS/QSLs intersections, we observe significant $\beta$ values (as well as slower wind speeds). This higher $\beta$ regime corresponds both to increased densities and lower magnetic field amplitude (in particular $B_r$). These extended large $\beta$ values also correspond to slightly weaker currents, as the HCS is thicker is these regions. In the fifth panel, we show the time averaged $\langle P_{cr} \rangle$ at $3 R_{\odot}$ between $t=60h$ and $t=180$h. As in Figure \ref{fig:stab_period}, gray/black colors indicate stable zones where magnetic tension contains the pressure gradient. Here again a structure is shaped by the network of quasi-separatrices. At most HCS/QSLs intersections, we observe very dark patches, which mark consistently stable zones over the period. Finally, the bottom panel displays the normalized histogram of every flux rope detected between $t=60h$ and $t=180h$ and $r=5 R_{\odot}$ and $10 R_{\odot}$ in the simulation of June 14. The detection algorithm is as follows: we trace magnetic field lines from seed points in a regular grid close to the HCS and between 5 and 10$R_{\odot}$, and we identify a flux rope when the number of sign changes of $B_r$ is larger than four along a given field line. This procedure can count multiple field lines which belong to the same flux rope at a given time, and may also catch the same propagating structure from one simulation output to the other (which are spaced by two hours). Hence, we normalize the distribution by its maximum to get a value between 0 and 1. This provides a general distribution of flux ropes occurrence in the simulation. As already noted in Figure \ref{fig:Fourier}, some longitudes are marked by more flux ropes than others, and we report the highest concentration of flux ropes between 150$^\circ$ and 270$^\circ$ longitude. 

These longitudes correspond to regions where the surrounding of the HCS, the heliospheric plasma sheet, is only marginally stable, i.e. white/clear regions in the fifth panel of Figure \ref{fig:squashing}. In contrast, we notice a very clear stable zone between 100 and 150 degrees of longitudes, where no flux rope are detected in the simulations. More precisely, we observe in the simulation lower temperatures and lower pressure gradients $\nabla p /p$ within QSLs. Typically, at $3R_{\odot}$, the plasma is around 1MK along the QSLs, while it is around 2MK in other open field regions. As the pressure gradients are lower, the regions around HCS/QSLs intersections are more stable. To test quantitatively this results, we compute in addition to the normalized statistics of flux ropes, a stability index, which is the sum on all latitudes of negative values of $\langle P_{\mathrm{cr}} \rangle$ normalized to fall back in the interval $[0,1]$. This index is shown in red in the last panel of Figure \ref{fig:squashing}. We then compute the Pearson correlation coefficient between the stability index and the flux ropes' count. We obtain a coefficient of $-0.42$, which indicates a statistically significant anti-correlation --no correlation is 0, while perfect (anti-)correlation is (-)1. We reproduce this analysis in Figure \ref{fig:squashing_app} for the simulation of June 1, and find similar results. 

\section{Summary and discussion}
\label{sec:ccl}

In this work, we have combined multi-spacecraft, multi-instrument analysis with 3D MHD simulations to investigate the origin of the  helical structures in the slow solar wind. We have taken advantage of the conjunction of the Parker Solar Probe fifth perihelion in June 2020 with the first measurements of Solar Orbiter and run 3D MHD simulations of the inner heliosphere at this period, from the chromosphere to $0.5$ AU. In contrast with a previous study, made at the first perihelion of PSP in November 2018, during a period of relatively low solar activity \citep{Reville2020ApJS}, several active regions have appeared during the month of June 2020. These rising phases of solar activity are associated with the formation of small coronal holes appearing at low latitudes in the active region belt \citep{Wang2010a}. As such, synoptic magnetic maps have evolved significantly over the whole month, and we found that one single simulation could not reproduce accurately the state of the inner heliosphere over the period, especially around the HCS. We looked for the optimal number and instances of synoptic magnetic field maps among the ADAPT database, and found that two maps, and thus two simulations, could render reasonably the magnetic sectors measured by Parker Solar Probe and Solar Orbiter. 

Reproducing the right magnetic sectors correctly is indeed necessary (but not sufficient) to identify and model numerically the sources of the solar wind. Comparing the in situ plasma data of PSP and Solar Orbiter with the simulations, we have shown a good overall agreement of the magnetic field amplitude, proton density and radial velocity. In the synthetic measurements made from the simulations, we observe periods of high and low variability in the density and velocity fields. These tend to occur when the probe is either crossing or close to the simulated HCS. Because both PSP and Solar Orbiter are confined to relatively low latitudes, they are bound to frequently cross the HCS and cruise in the heliospheric plasma sheet, and although there are many additional sources of variability in the actual data (Alfvén waves, switchbacks), we do find a good correspondence between regions of strong variability in the simulations, and region of low cross-helicity and high plasma beta, i.e. when the spacecraft is close to the HCS. We analyzed further the data from PSP to identify individual HCS crossings and flux rope events, using in particular the pitch-angle distributions of suprathermal electrons (see Figure \ref{fig:beta_comp} and Figure \ref{fig:fr_psp}).

Our simulations also reproduce flux ropes, confined to the HCS. They are created at the tip of helmet streamers by a succession of instabilities. First, a pressure driven instability extends and thins the current sheet. Then, a tearing mode disrupts the sheet, triggering reconnection following a guide field in the super-alfvénic regime. This process is very similar to the one described in 2.5D by \citet{Reville2020ApJL}, in which we clearly identified the characteristics of the resistive tearing mode as the source of chains of plasmoids released in the slow wind. However, as shown in \citet{Reville2020ApJL}, the complete tearing process cannot be fully characterized in low Lundquist numbers regime. In this work, we run ideal MHD simulations and the resistivity is set by the numerical grid and scheme. We estimate the Lundquist number to be around $500-1000$ close to the reconnection region of the simulations. The threshold to trigger the tearing mode is thus lower in these 3D simulations than in the 2.5D configuration studied in \citet{Reville2020ApJL}. Recent work has shown that the presence of an inhomogeneous guide field could lead to an enhanced growth rate of the tearing instability \citep{Lotfi2021}.  More generally, the complex configurations induced by the realistic magnetic fields must change the picture in comparison with the purely axisymmetric case: however, since coronal Lundquist numbers should be substantially greater than in our simulations and of order $S \sim 10^{14}$, one expects tearing instabilities to occur naturally in the forming HCS.

The simulations confirm the ability of sequential magnetic reconnection at the tip of helmet streamers to produce flux ropes in the HCS \citep[see also][for another 3D study]{HigginsonLynch2018}. These flux ropes are first connected on both sides to the Sun, but can also reconnect later in the open solar wind, which is consistent with typical observed electrons pitch-angle distributions. The 3D simulations also reveal a long periodicity of the flux rope release, about 20h, corresponding to the formation and re-formation of the HCS after the pressure driven instability. Following \citet{Peterson2021}, we have computed the characteristic timescale of this process, and we recover the long timescale (20h-30h) observed in 2.5D and 3D simulations. Note that, in \citet{Peterson2021}, the heuristics yields shorter frequencies, directly identified to the main peak of observed periodic density perturbations of the solar wind, around 90 minutes \citep{Viall2015}. However, their computation does not include the gravity pull, which leads to an interesting difference between the latter and the present work. Removing the gravitational acceleration in equation \ref{eq:taucr}, we find an average periodicity of the order of $3h$. But, we also find that, without gravity, the whole inner corona is unstable to pressure gradient forces, which means that gravity does play an important role in the force balance in these regions. Hence, while they claim that the specific details of the tearing mode do not matter to reach agreement with observations, we argue that the fastest growing mode of the ideal tearing is necessary to go down to the hour-long timescales, as shown in \citet{Reville2020ApJL}.

Finally, we observe some structure in the longitudinal distribution of flux ropes events. The global MHD solution is shaped by the topology of the magnetic field, and in particular the structure of separatrices and quasi-separatrices, which can be easily identified with the squashing factor $Q$. We obtain slower wind, higher beta, lower temperatures and lower currents at some intersections of the QSLs with the HCS. These intersections are consequently less prone to the pressure driven instability, and they determine the end of the reconnection line and thus the extent of the flux rope. It is not trivial to generalize this result to high S regimes, as we can expect the current sheets to always be unstable to the tearing mode for low enough finite resistivity \citep{Biskamp1986}. Yet, the pressure instability analysis should be robust in any resistive regime. Further studies are necessary to characterize the structure of flux ropes in 3D realistic configurations and higher S regimes. This is crucial for a better understanding of the connectivity in the HCS, which can be completely different from what is predicted by static models. 

\section{Acknowledgement}
This research was funded by the ERC SLOW{\_}\,SOURCE project (SLOW{\_}\,SOURCE - DLV-819189). VR, MV, CS acknowledge discussions within the HERMES DRIVE Science center team. The authors are grateful to A. Mignone and the PLUTO development team. Simulations were performed on the Jean-Zay supercomputer (IDRIS), through GENCI HPC allocations grants A0090410293 and A0110410293. Parker Solar Probe was designed, built, and is now operated by the Johns Hopkins Applied Physics Laboratory as part of NASA's Living with a Star (LWS) program. The SWEAP and FIELDS investigations are supported by the PSP mission under NASA contract NNN06AA01C. Solar Orbiter is a space mission of international collaboration between ESA and NASA, operated by ESA. Solar Orbiter Solar Wind Analyser (SWA) data are derived from scientific sensors which have been designed and created, and are operated under funding provided in numerous contracts from the UK Space Agency (UKSA), the UK Science and Technology Facilities Council (STFC), the Agenzia Spaziale Italiana (ASI), the Centre National d’Études Spatiales (CNES, France), the Centre National de la Recherche Scientifique (CNRS, France), the Czech contribution to the ESA PRODEX programme and NASA. Solar Orbiter SWA work at UCL/MSSL is currently funded under STFC grants ST/T001356/1 and ST/S000240/1. Solar Orbiter magnetometer operations are funded by the UK Space Agency (grant ST/T001062/1). This work utilizes data produced collaboratively between AFRL/ADAPT and NSO/NISP. Data analysis was performed with the help of the AMDA science analysis system provided by the Centre de Données de la Physique des Plasmas (CDPP) supported by CNRS, CNES, Observatoire de Paris and Université Paul Sabatier, Toulouse.  Work at IRAP and LAB was performed with the support of CNRS and CNES. ASB and AS were supported by a CNES Solar Orbiter grant and PNST. SP acknowledges the funding by CNES through the MEDOC data and operations center. This study has made use of the NASA Astrophysics Data System.

\bibliographystyle{aa}
\bibliography{./biblio.bib}

\begin{appendix}
\section{June 1 simulation}
\label{app:June1}

\begin{figure}
\includegraphics[width=3.5in]{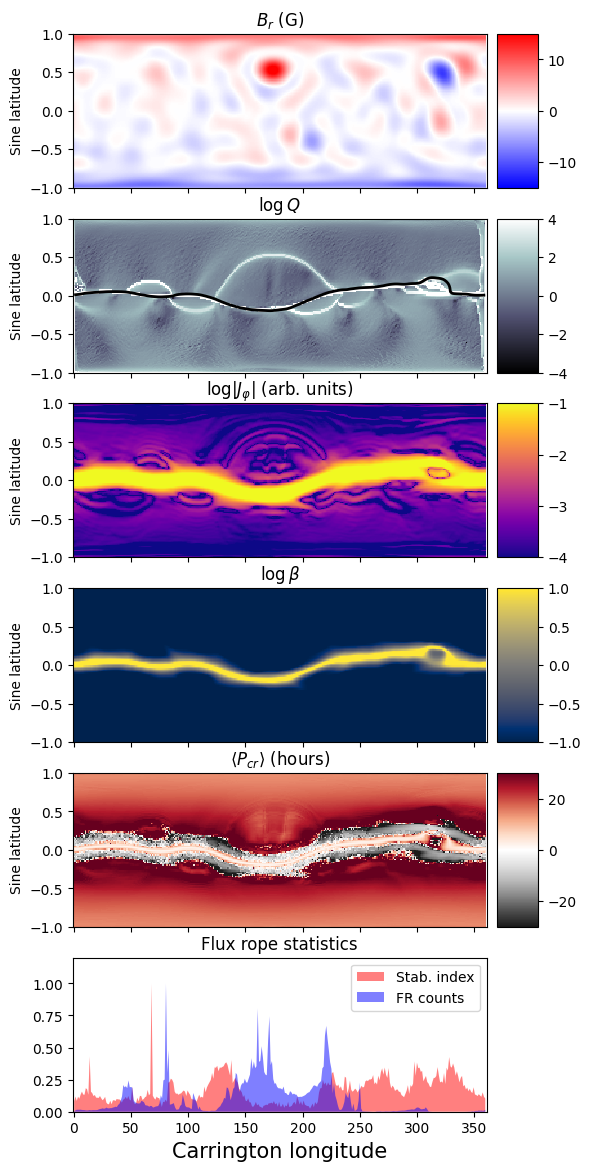}
\caption{Same as Figure \ref{fig:squashing} for the June 1 simulation.}
\label{fig:squashing_app}
\end{figure}

Figure \ref{fig:squashing_app}, repeats the same analysis as done for Figure \ref{fig:squashing} for the simulation of June 1. At this time, the lower hemisphere active region (AR) has not appeared on the solar disk (it will on June 3 at around 100° of longitude). The AR around 320 degrees appears stronger compared to the June 14 map. The current sheet shows a significantly weaker warp and a simpler QSLs and current structures. However, at high longitudes, we do observe a complex structure of QSLs and a thickening of the HCS. Stable regions with few flux ropes are thus located at longitudes between 250 and 330 degrees of longitudes. The Pearson correlation coefficient between the flux ropes counts and the normalized stability curve is $-0.3$.

\section{Flux rope and HCS crossings identification}
\label{app:FRtable}

In this work, we visually inspected PSP data from May 28, 2020, to June 29, 2020, in order to identify structures linked to HCS dynamics such as heliospheric current sheet crossings (HCSs), flux ropes (FRs) and coronal mass ejections (CMEs). Although we intend to be as objective as possible, this selection relies on visual inspection. For reproducibility purposes, we provide the list of the identified structures in table \ref{tab:HCS_dynamics}. The HCS crossings were identified by a reversal of the radial component of the magnetic field associated with a change in the pitch-angle distribution (PAD) of the suprathermal electron population (strahl) consistent with its outward propagation. We identify flux ropes based on their classical signatures, namely an increase in the magnetic pressure and a bipolar signature in one of the components of the magnetic field. The boundaries of the events were defined based on variations in the profiles of the magnetic and plasma parameters.

\begin{table}[h]
    \centering
    \begin{tabular}{c|c c|l}
\# & Start & End & Description\\
\hline
\hline
&&&\\
1	& 2020-05-27 17h00 & 2020-05-28 04h00 & HCS with FRs \\
2	& 2020-05-28 08h58 & 2020-05-28 14h55& CME\\
3	& 2020-05-29 21h40 & 2020-05-30 04h31& CMEs \\
4	& 2020-05-31 00h00 & 2020-05-31 07h00& HCS with FRs\\
5	& 2020-05-31 13h04 & 2020-05-31 03h19& PHCS with CMEs/FR\\
6	& 2020-06-01 10h20 & 2020-06-01 12h16& PHCS\\
7	& 2020-06-01 12h52 & 2020-06-01 13h02& HCSs \\
8	& 2020-06-01 14h36 & 2020-06-01 16h10& CME\\
9	& 2020-06-01 19h21 & 2020-06-01 21h36& HCSs and FR\\
10	& 2020-06-02 06h41 & 2020-06-02 09h19& CMEs with HCS\\
11	& 2020-06-04 03h30 & 2020-06-04 06h05& PHCS\\
12	& 2020-06-07 07h07 & 2020-06-07 08h50& CME\\
13	& 2020-06-07 11h19 & 2020-06-07 12h34& PHCS\\
14	& 2020-06-07 20h16 & 2020-06-07 21h09& PHCS with FRs\\
15	& 2020-06-07 23h30 & 2020-06-08 12h30& HCS with FRs\\
16	& 2020-06-08 15h30 & 2020-06-09 01h40& HCS with FRs\\
17	& 2020-06-12 01h30 & 2020-06-12 07h50& CME\\
18	& 2020-06-16 16h00 & 2020-06-17 14h00& PHCS with FRs\\
19	& 2020-06-18 08h00 & 2020-06-18 12h30& HCSs with FRs\\
20	& 2020-06-18 18h40 & 2020-06-19 00h30& HCSs with FRs\\
21	& 2020-06-19 11h10 & 2020-06-20 12h00& HCSs with FRs\\
22	& 2020-06-21 06h00 & 2020-06-21 14h00& HCSs with FRs\\
23	& 2020-06-23 07h00 & 2020-06-23 17h00& HCSs with FRs\\
24	& 2020-06-24 12h00 & 2020-06-25 03h00& HCSs with FRs\\
25	& 2020-06-25 11h00 & 2020-06-27 12h00& CME (Fig. \ref{fig:app_cme})\\
    \end{tabular}
    \label{tab:HCS_dynamics}
    \caption{Identified (partial) Heliospheric Current Sheet (PHCS/HCS) crossings, Coronal Mass Ejections (CMEs), and Flux Ropes (FR) in PSP data during the month going from May 28, 2020, to June 29, 2020}
\end{table}

\begin{figure}
    \centering
    \includegraphics[width=3.5in]{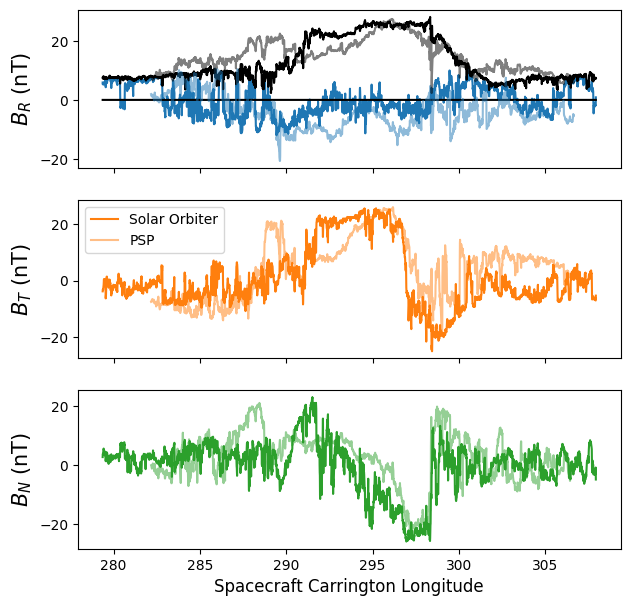}
    \caption{Two CME event discovered in the vicinity of the HCS by Solar Orbiter and PSP one month apart. The Carrington longitude of PSP is shifted by 17 degrees to superpose the structures. The total field $||B||$ is shown in black (Solar Orbiter)/gray (Parker Solar Probe) in the top panel.}
    \label{fig:app_cme}
\end{figure}

In Figure \ref{fig:beta_comp}, we overlay in red all HCS, PHCS and FR events of this table. Because the relationship between CMEs and the HCSs is out of the scope of this paper, we have removed all pure CME events of Figure \ref{fig:beta_comp}. However, we discovered an interestingly very similar CME event observed one month apart by both PSP and Solar Orbiter. The CME is reported as the 25th event of Table \ref{tab:HCS_dynamics}, measured between June 25 and June 27. Solar Orbiter measured another CME between May 28 and May 30. In Figure \ref{fig:app_cme}, we show the magnetic field structure of the two events as a function of the Carrington longitude of each spacecraft. We added 17 degrees to the longitude of PSP to superpose the structures. The amplitude, orientation and overall structures of the two CMEs are very close, which suggests that they might come from a similar origin.

\end{appendix}
\end{document}